The Impact of Banking Competition on Interest Rates for Household

Consumption Loans in the Euro Area

Alexander Rom

University of San Francisco

May 17, 2024

**Abstract:**

This paper investigates the impact of banking competition on interest rates for household consumption loans in the Euro Area from 2014 to 2020. Utilizing a panel data regression approach, we analyze how various factors, including local banking competition, influence the interest rates set by banks across 13 Euro-area countries. Our key independent variable, local banking competition, is measured by the number of commercial bank branches per 100,000 adults. Control variables include the ECB interest rate, euro exchange rate, real GDP growth rate, inflation rate, unemployment rate, bank business volumes, and country risk. We address potential endogeneity and heterogeneity biases and employ both Fixed Effects and Hausman-Taylor models to ensure robust results. Our findings indicate that higher local banking competition is associated with a slight increase in interest rates for household loans. Additionally, factors such as ECB interest rate, country risk, and euro appreciation significantly affect interest rates. The results offer insights into how competitive dynamics in the banking sector influence borrowing costs for households, providing valuable implications for policymakers and financial institutions in the Euro Area.

# 1 Introduction

If we look at the differences in interest rates for any bank loan between Euro-area countries in any given year we will notice significant differences between them from country to country. For example, in March of 2024 (image 1.1), the Euro-area average was 7.82% for household loans for consumption. While some countries, such as Malta, had interest rates of around 4.23%, others, like Estonia, were experiencing rates as high as 12.93%. This paper attempts to investigate what factors play a role in such a large variation in rates across countries by addressing two key quotations.

- What factors influence the interest rates on bank loans for households in Euro-area countries?

- Is there any relationship between interest rates and banking competition in each country?

Understanding answers to these quotations about interest rates is crucial for both policymakers and financial institutions, as these rates directly affect consumer borrowing costs and overall economic activity. We employ panel data regression techniques, and use both Fixed Effects and Hausman-Taylor models, to address potential endogeneity and heterogeneity biases in the analysis to find the most accurate estimations for the effects. Our findings reveal a nuanced relationship between banking competition and interest rates, with higher competition associated with slight increases in interest rates for household loans. Additionally, the study highlights the significant impact of other macroeconomic factors on interest rates, providing a comprehensive understanding of the determinants of borrowing costs in the Euro Area.

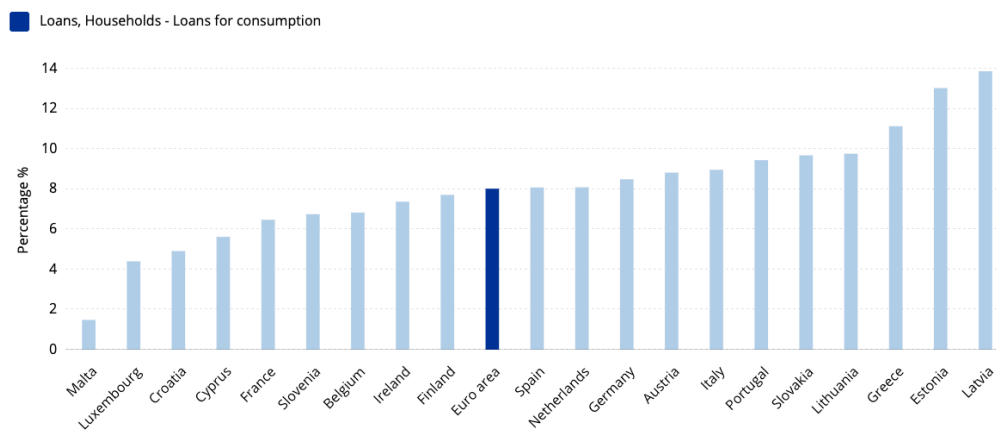

*(1.1)*

# 2    Data Description

This study uses a balanced panel dataset, over the period from 2011 to 2023, resulting in a time dimension (t) of 13 years and covering 13 Euro-area countries (i): Austria, Belgium, Finland, France, Germany, Greece, Ireland, Italy, Lithuania, Luxembourg, Portugal, Slovenia, Spain. Next are listed all variables used in the model.

**Outcome variable:**

**IR**(it) = bank interest rate for household consumption loans (%)

**Treatment variable:**

**BComp**(it) = commercial bank branches (per 100,000 adults)

**Control variables:**

**ECB_rate**(t) = ECB interest rate;

**EXCH_rate**(t) = Euro exchange rate as Real Broad Effective Exchange Rate for Euro Area;

**GDP**(it) = Real GDP growth rate;

**INFL**(it) = HICP - inflation rate;

**U**(it) = Unemployment rate;

**ALM**(it) = Bank business volumes - loans to households for consumption (new business)

**CR**(it) = Country risk as Difference between 10-year government bond(i) and the 10-year rate for Eurobond

**LITH_2014** = dummy variable for Lithuania in 2014 equaling 1

**INFL_sq**(it) = squared INFL, there is non-liber relation between IR and INFL so transfrontion is needed, after testing squared transformation prefremed the best

**BComp_trend**(t)  = component of BComp that is explained by the linear time trend, holding country-specific effects constant

**Covid_2020** = dummy variable for COVID-19, assigned a value of 1 for each country in the year 2020

**Descriptive Statistics**

| Variable | Obs | Mean | Std. Dev. | Min | Max |
|----------|-----|------|-----------|-----|-----|
| IR | 105 | 6.818 | 3.301 | 1.986 | 16.248 |
| BComp | 105 | 30.89 | 18.07 | 4.02 | 79.57 |
| BComp trend | 105 | 30.89 | 3.118 | 26.236 | 35.545 |
| CR | 91 | .215 | 1.565 | -1.122 | 8.401 |
| ECB rate | 105 | .031 | .055 | 0 | .16 |
| EXCH rate | 105 | 98.944 | 2.772 | 94.66 | 103.71 |
| GDP | 105 | 1.747 | 4.122 | -11.2 | 24.5 |
| INFL | 105 | .73 | 1.037 | -1.5 | 3.7 |
| INFL sq | 105 | 1.6 | 2.484 | 0 | 13.69 |
| LITH 2014 | 105 | .01 | .098 | 0 | 1 |
| U | 105 | 9.416 | 5.092 | 3 | 26.6 |
| covid 2020 | 105 | .143 | .352 | 0 | 1 |
| ALM | 105 | 16139.981 | 28315.588 | 39 | 110866 |

*(2.1)*

From the descriptive statistics table (2.1), we can see that we have 91 observations available for our models, as the CR variable does not include data for Estonia and Cyprus. However, other variables do have observations from these countries, bringing the total number of observations to 105. Additionally, we see a large variation in minimum to maximum values for most of the variables.

If we look at the scatterplot (2.2) with BComp on the x-axis and IR on the y-axis we will see that there visually negative relationship (corr(BComp, IR = -0.4273). It also might seem that there is a non-linear relationship, however, the models with squared or lagged BComp variables had lower significance so the decision was made to not do any trasfrontions. Also, we can see (2.3) an overall decreasing trend for all countries in BComp (overall decrease in BComp from 35.54 to 26.23, total change 9.31 across all countries) suggesting that with time accessibility to traditional banking is falling.

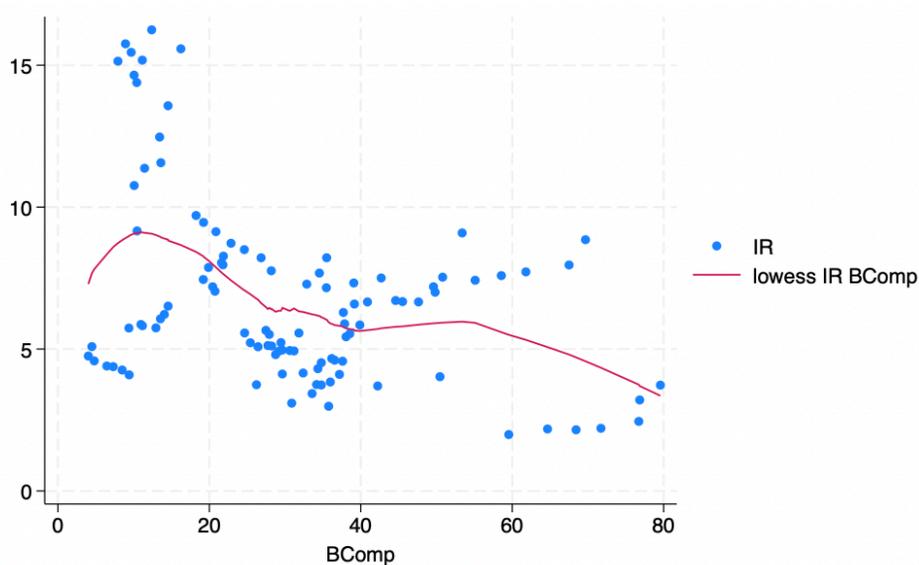

*(2.2)*

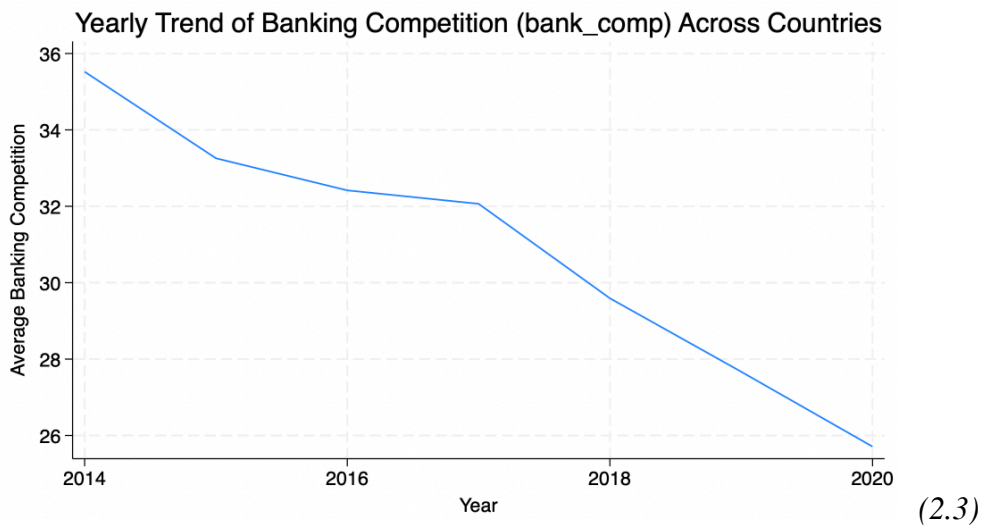

*(2.3)*

# 3    Empirical Model and Identification

We can assume to have several identification issues as we are working with panel data, Most likely endogeneity bias is present as COV(CR(it); u(it)) ≠ 0, same for GDP(it) and U(it). We know from theory that these variables are likely to be subject to random shocks that we can not take into account such as war or natural disasters that have significant effect on macroeconomic indicators. Also, heterogeneity bias is likely to be an issue with our model suggesting that COV(BComp(it); a(i)) ≠ 0), and the same for GDP(it) & CR(it). Time invariant error will correlate with banking competition as we know that factors like banking culture that was established by centuries or decades play a role in determining banking competition. Additionally, we have mentioned before the issue of missing observation for country risk that reduces the number of observations in our model and causes it to be less accurate. At last, our data does not cover the full economic cycle, from one recession to another so we can not have a full understanding of how these variables perform in different time conditions. Image (3.1) shows that in the time period covered by our data from 2014 to 2020 Euro-area countries experienced a time of economic growth and decreasing interest rates by the central bank, only in 2020 we had negative growth which was caused by the pandemic and lockdown of the economy.

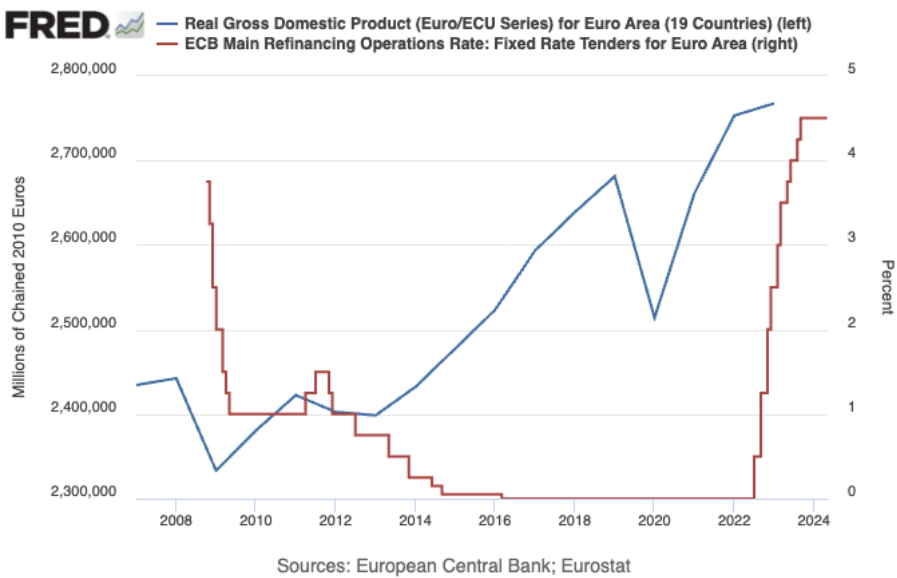

*(3.1)*

Moving to the empirical model, we use a simple OLS regression incorporating all our variables.

**Linear regression**

| IR | Coef. | St.Err. | t-value | p-value | [95% Conf | Interval] | Sig |
|---|---|---|---|---|---|---|---|
| BComp | -.05 | .012 | -4.16 | 0 | -.074 | -.026 | *** |
| BComp_trend | -.138 | .202 | -0.68 | .496 | -.54 | .264 | |
| GDP | .111 | .063 | 1.75 | .084 | -.015 | .237 | * |
| INFL | -.558 | .495 | -1.13 | .263 | -1.543 | .428 | |
| INFL_sq | .362 | .175 | 2.06 | .042 | .013 | .711 | ** |
| ALM | 0 | 0 | -0.62 | .534 | 0 | 0 | |
| U | .221 | .069 | 3.21 | .002 | .084 | .358 | *** |
| EXCH_rate | -.064 | .142 | -0.45 | .651 | -.346 | .218 | |
| ECB_rate | 7.075 | 11.621 | 0.61 | .544 | -16.06 | 30.21 | |
| CR | -.036 | .213 | -0.17 | .866 | -.461 | .389 | |
| LITH_2014 | 8.191 | 1.98 | 4.14 | 0 | 4.249 | 12.133 | *** |
| covid_2020 | .004 | 1.103 | 0.00 | .997 | -2.192 | 2.201 | |
| Constant | 15.992 | 18.265 | 0.88 | .384 | -20.372 | 52.355 | |

| | | | | |
|---|---|---|---|---|
| Mean dependent var | 6.378 | SD dependent var | 2.486 |
| R-squared | 0.504 | Number of obs | 91 |
| F-test | 6.596 | Prob > F | 0.000 |
| Akaike crit. (AIC) | 385.216 | Bayesian crit. (BIC) | 417.857 |

*** $p<.01$, ** $p<.05$, * $p<.1$

Based on the linear regression model results, we can draw several conclusions regarding the significant variables affecting interest rates for household consumption loans. Banking competition (BComp), has a high significance with a p-value of 0.000. Suggesting that an increase in banking competition is associated with a decrease in interest rates, with a one-unit increase in competition leading to a 0.05% decrease in interest rates, while holding other variables constant. The unemployment rate (U) is another significant variable, with a p-value of 0.002. Higher unemployment rates are associated with higher interest rates for household consumption loans, as a one percentage point increase in unemployment leads to a 0.221% increase in interest rates, holding other variables constant. Additionally, the variable representing Lithuania's adoption of the euro in 2015 (LITH_2014) is highly significant with a p-value of 0.000, suggesting that this event resulted in a substantial increase in interest rates, specifically an 8.191% rise, holding other variables constant. Finally, our last significant variable INFL_sq shows significance with a p-value of 0.042, indicating a nonlinear relationship where interest rates rise at an increasing rate as inflation increases. If we look at the overall effect of inflation,

$$-0.5458089 \times \text{INFL} + 0.3500378 \times \text{INFL}^2 = 0$$

$$\text{INFL} = \frac{0.5458089}{0.3500378} \approx 1.559$$

The calculation shows that the effect of inflation on interest rates starts to become positive (as opposed to negative with low inflation) when the inflation rate exceeds approximately 1.559%, holding other variables constant.

## 4     Results

**A.  Fixed-Effects & Random-Effects**

For both fixed-effects (FE) and random-effects, we apply robust standard errors that are designed to be valid even when heteroskedasticity is present. Additionally, this approach helped to increase the significance of all of the variables in the model.

**RE (Robust)**

**Regression results**

| IR | Coef. | St.Err. | t-value | p-value | [95% Conf | Interval] | Sig |
|---|---|---|---|---|---|---|---|
| BComp | -.005 | .03 | -0.17 | .863 | -.065 | .054 | |
| BComp_trend | -.03 | .084 | -0.36 | .719 | -.196 | .135 | |
| GDP | -.01 | .027 | -0.36 | .716 | -.064 | .044 | |
| INFL | -.073 | .344 | -0.21 | .833 | -.747 | .602 | |
| INFL_sq | -.022 | .102 | -0.22 | .827 | -.223 | .178 | |
| ALM | 0 | 0 | -0.99 | .324 | 0 | 0 | |
| U | .068 | .059 | 1.14 | .254 | -.049 | .184 | |
| CR | -.254 | .06 | -4.26 | 0 | -.371 | -.137 | *** |
| EXCH_rate | -.029 | .03 | -0.98 | .325 | -.088 | .029 | |
| ECB_rate | 3.877 | 2.721 | 1.42 | .154 | -1.456 | 9.211 | |
| LITH_2014 | 3.671 | .297 | 12.38 | 0 | 3.089 | 4.252 | *** |
| covid_2020 | -.614 | .411 | -1.49 | .135 | -1.42 | .192 | |
| Constant | 10.084 | 3.538 | 2.85 | .004 | 3.149 | 17.019 | *** |
| Mean dependent var | | 6.378 | SD dependent var | | | 2.486 | |
| Overall r-squared | | 0.164 | Number of obs | | | 91 | |
| Chi-square | | 974481.658 | Prob > chi2 | | | 0.000 | |
| R-squared within | | 0.542 | R-squared between | | | 0.151 | |

*** p<.01, ** p<.05, * p<.1

The model explains about 54.24% of the variation in interest rates within countries and about 16.41% overall. The Wald chi-squared statistic is highly significant, indicating that the model as a whole is statistically significant.

We will focus on a few key variables to learn how they change across models. The coefficient for banking competition (BComp) is -0.005128 with a p-value of 0.863, indicating that the variable is not statistically significant. However, the result is consistent with our OLS models suggesting that higher BComp values have a negative effect on the IR. In contrast, country risk (CR) which was not significant in our OLS model now shows a highly significant effect on interest rates, with a coefficient of -2.55976 and a p-value of 0.000. This negative relationship implies that higher country risk is associated with lower interest rates for household loans. Specifically, a one-unit increase in country risk leads to a decrease in interest rates by 2.56%, holding other variables constant. This result is counter-intuitive and goes against the theory, but it will be discussed in full in the conclusion. Finally, the adoption of the euro by Lithuania in 2015 (LITH_2014) is found to have a significant positive impact on interest rates as in our OLS. The coefficient for LITH_2014 is 3.670642 with a p-value of 0.000, indicating that this event led to a substantial increase in interest rates by about 3.67%, however was not as big as we had in the OLS model where the coefficient was more than double this size.

**FE (robust)**

**Regression results**

| IR | Coef. | St.Err. | t-value | p-value | [95% Conf | Interval] | Sig |
|---|---|---|---|---|---|---|---|
| BComp | .055 | .025 | 2.20 | .048 | .001 | .109 | ** |
| BComp_trend | -.04 | .089 | -0.45 | .66 | -.234 | .154 | |
| GDP | -.011 | .026 | -0.41 | .689 | -.067 | .046 | |
| INFL | -.008 | .332 | -0.02 | .981 | -.732 | .715 | |
| INFL_sq | -.05 | .089 | -0.56 | .589 | -.244 | .145 | |
| ALM | 0 | 0 | -0.44 | .67 | 0 | 0 | |
| U | -.033 | .087 | -0.38 | .71 | -.224 | .157 | |
| CR | -.235 | .058 | -4.05 | .002 | -.361 | -.109 | *** |
| EXCH_rate | -.044 | .038 | -1.16 | .268 | -.125 | .038 | |
| ECB_rate | 4.743 | 2.878 | 1.65 | .125 | -1.528 | 11.014 | |
| LITH_2014 | 3.635 | .265 | 13.73 | 0 | 3.058 | 4.212 | *** |
| covid_2020 | -.395 | .425 | -0.93 | .371 | -1.321 | .531 | |
| Constant | 10.676 | 4.066 | 2.63 | .022 | 1.818 | 19.534 | ** |

| | | | | | | |
|---|---|---|---|---|---|---|
| Mean dependent var | | 6.378 | SD dependent var | | | 2.486 |
| R-squared | | 0.591 | Number of obs | | | 91 |
| F-test | | . | Prob > F | | | . |
| Akaike crit. (AIC) | | 139.052 | Bayesian crit. (BIC) | | | 166.672 |

*** p<.01, ** p<.05, * p<.1

The model explains about 59.06% of the variation in interest rates within countries and about 5.87% overall. This indicates that the fixed effects model captures a substantial portion of the variation within countries but not as much overall.

Coefcints for our significant variables LITH_2014 and CR stayed almost the same so we will not discuss them once more. However, for BComp we see a significant change in the coefficient. The FE regression model indicates that banking competition (BComp) has a significant positive impact on interest rates for household consumption loans in the Euro Area. With a coefficient of 0.0548417 (opposite coefficient compared to our RE model) and a p-value of 0.048, the results show that a one-unit increase in BComp is associated with a 0.0548% increase in interest rates, holding other variables constant.

**Hausman Test**

We use the Hausman test to compare the coefficients from the FE model and the RE model to determine which model is appropriate. The null hypothesis (H0) is that the difference in coefficients is not systematic, implying that the RE model is preferred. The alternative hypothesis (Ha) is that the difference is systematic, suggesting that the FE model is more appropriate.

**Hausman (1978) specification test**

| | Coef. |
|---|---|
| Chi-square test value | 25.196 |
| P-value | .001 |

Since the p-value is 0.001, which is below 0.05, we reject the null hypothesis. Meaning that there is a significant difference between the coefficients of the FE and RE models, and thus FE model is appropriate.

## B. <u>Hausman-Taylor model</u>

For our Hausman-Taylor model, we get a highly significant Wald chi-squared statistic (equaling 96.14) and p-value = 0.000 which indicates that the model as a whole is statistically significant.

```
------------------------------------------------------------------------------
          IR │ Coefficient  Std. err.      z    P>|z|     [95% conf. interval]
-------------┼----------------------------------------------------------------
TVexogenous  │
  BComp_trend│  -.0327438    .067447    -0.49   0.627    -.1649374    .0994498
          ALM│  -9.77e-06    .0000135   -0.72   0.469    -.0000362    .0000167
     ECB_rate│   4.719053   3.416158     1.38   0.167    -1.976493   11.4146
    LITH_2014│   3.666819    .6308949    5.81   0.000     2.430287    4.90335
   covid_2020│  -.3913841    .3590303   -1.09   0.276    -1.095071    .3123025
TVendogenous │
          GDP│  -.0094527    .0264832   -0.36   0.721    -.0613589    .0424535
           CR│  -.2364144    .0969651   -2.44   0.015    -.4264625   -.0463662
         INFL│  -.0144541    .1644563   -0.09   0.930    -.3367826    .3078743
      INFL_sq│  -.0444137    .05665     -0.78   0.433    -.1554457    .0666182
    EXCH_rate│  -.0433289    .0435436   -1.00   0.320    -.1286728    .042015
            U│  -.0318934    .0622942   -0.51   0.609    -.1539877    .0902009
        BComp│   .0468804    .0234237    2.00   0.045     .0009707    .09279
TIexogenous  │
   country_id│   .1142285    .211693     0.54   0.589    -.3006821    .5291392

        _cons│   9.679687   5.843526     1.66   0.098    -1.773414   21.13279
-------------┼----------------------------------------------------------------
      sigma_u│  2.9202694
      sigma_e│   .49722151
          rho│   .97182638   (fraction of variance due to u_i)
------------------------------------------------------------------------------
```

The results of this model are very similar to the FE model outcome as the same coefficients stayed significant and the direction of their effect stayed the same. Banking competition (BComp) is statistically significant with a coefficient of 0.0468804 and a p-value of 0.045. Country risk (CR) also shows a significant negative impact on interest rates, with a coefficient of -0.2364144 and a p-value of 0.015. And Lithuania's euro adoption in 2015 (LITH_2014) is highly significant with a coefficient of 3.666819 and a p-value of 0.000. In more detail, the outcomes of the coefficient will be discussed in the conclusion section.

# 5    Conclusion

This study investigates the impact of banking competition, and other macroeconomic indicators on interest rates for household consumption loans in the Euro Area from 2014 to 2020, employing panel data regression techniques including fixed effects, random effects, and the Hausman-Taylor model. The results from the Hausman-Taylor model provide robust insights into the determinants of interest rates and specifically how our key variable of banking competition is related to interest rates.

Country risk (CR) is a critical factor influencing interest rates. The Hausman-Taylor model indicates a significant negative relationship between country risk and interest rates, implying that higher economic and political instability leads to lower interest rates. Unxepetcd negative relation can be explained by possible reductions in borrowing costs, as banks lower rates to attract more secure loans in riskier environments. Additionally,  in response to increased country risk, which might include economic instability, political uncertainty, or financial crises, a central bank might lower benchmark interest rates to stimulate economic activity.

The adoption of the euro by Lithuania in 2015 (LITH_2014) also shows a significant impact on interest rates, with the model indicating a substantial increase following the adoption. This result highlights the significant economic adjustments associated with major monetary policy changes and the effects of adopting a more widely used currency.

Additionally, the Hausman-Taylor model suggests (with low significance for these coefficients) that when Euro (EXCH_rate) appreciation by 1 unit (mean 98.94) causes interest rates to decrease by 0.043%. We also can assume that a 1% increase in the ECB rate (ECB_rate) causes a 4.72% increase in interest rates for consumer loans in Euro-area countries. Covid_2020: The COVID-19 pandemic (covid_2020) variable indicates a negative impact on interest rates, with the model showing a decrease of approximately 0.39%. All these outcomes are theoretically sound and can be explained by monetary economics.

At last, the analysis of the model reveals that banking competition (BComp) has a statistically significant positive impact on interest rates. Specifically, an increase in the number of commercial bank branches per 100,000 adults is associated with a rise in interest rates for household loans by approximately 0.0469%. This finding might suggest that higher competition among banks may lead to increased operational costs and thus higher interest rates for consumer loans. The reason our result for this variable may not align with theoretical expectations could be that our BComp variable is not the most accurate measure of banking competition. The number of commercial bank branches per 100,000 adults measures accessibility to traditional banking services, such as the physical location of banks. However, it is possible that all these branches are owned by a single bank, which means our variable may fail to accurately capture true banking competition. Thus, for future studies on banking competition, it would be more accurate to find or create variables that better capture banking competition, such as the number of licensed banking institutions per 100,000 people.

  Even though this project fails to determine the exact relationship between interest rates and banking competition in each country, we are still able to establish some factors that influence these consumer loan rates. Overall, this study provides valuable insights into the determinants of interest rates for household consumption loans in the Euro Area

# 7    Appendix

```
// convert montly to annual data
gen year_num = real(substr(timeperiod, 1, 4))
egen annual_mean_bank_comp = mean(bank_comp), by(year_num country_name)
bysort year_num country_name: keep if _n == 1 // dorop duplicate

merge 1:m country year using "/Users/alexanderrom/Desktop/EUR/org copy/GDP.dta"
merge 1:m country year using "/Users/alexanderrom/Desktop/EUR/org copy/inflation.dta"
merge 1:m country year using "/Users/alexanderrom/Desktop/EUR/org copy/loan_amount.dta"
merge 1:m country year using "/Users/alexanderrom/Desktop/EUR/org copy/unemployment.dta"
merge 1:m country year using "/Users/alexanderrom/Desktop/EUR/org copy/loan_rates.dta"

replace Rate = subinstr(Rate, ",", ".", .)
destring Rate, replace
replace EU_bond_rate = round(EU_bond_rate, 0.001)

////////////////////////////////////////////////////////////////////

//Generate country_id:
egen country_id = group(country_name), label
// GEn time id
```

```
summarize year, meanonly
local min_year = r(min)
gen time_id = year - `min_year`

//////////////////////////////////////////////////////////////////////////

// bank comp through tiem
regress BComp year
generate trend = _b[_cons] + _b[year]*year
// Country-specific Trends
xtset country_id year
xtreg BComp year, fe
predict trend_country, xb
twoway (line BComp year if country_id == 1, sort) (line trend_country year if country_id == 1,
sort)
// include a separate intercept for each country, effectively controlling for all time-invariant
differences across countries. This means any unobserved variable that does not change over time
within each country but varies between countries will not bias the estimated effect of time.
gen BComp_LITH2014 = BComp * LITH_2014
// graphs
scatter IR BComp || lowess IR BComp
//////////////////////////////////////////////////////////////////////////

//Generating the yearly average of bank_comp
collapse (mean) BComp, by(year)
//Sorting the data by year to ensure the plot follows chronological order
sort year
//Finding the maximum value of bank_comp to set the y-axis scale dynamically
summarize BComp, detail
local max_BComp = r(max)
* Creating a line plot of the average bank_comp over the years with a y-axis ranging from 0 to
max value
twoway (line bank_comp year), title("Yearly Trend of Banking Competition (bank_comp)
Across Countries") \
xtitle("Year") ytitle("Average Banking Competition") yscale(range(0 `max_bank_comp`))
legend(off) graphregion(color(white) lcolor(black))
```

```
/////////////////////////////////////////////////////////////////////////
// OLS model
reg IR BComp GDP INFL INFL_sq ALM U EXCH_rate ECB_rate CR LITH_2014 covid_2020

// IV model
gen BComp_lag1 = L.BComp
ivregress 2sls  IR (BComp= BComp_trend) GDP INFL INFL_sq ALM U EXCH_rate ECB_rate
CR LITH_2014 covid_2020

// 2SLS RE and FE
xtivreg IR (BComp= BComp_trend) GDP INFL INFL_sq ALM U EXCH_rate ECB_rate CR
LITH_2014 covid_2020, re vce(robust)
xtivreg IR (BComp = BComp_trend) GDP INFL INFL_sq ALM U EXCH_rate ECB_rate
LITH_2014 covid_2020, fe vce(robust)

// RE and FE
xtreg IR BComp BComp_trend GDP INFL INFL_sq ALM U CR EXCH_rate ECB_rate
LITH_2014 covid_2020, fe robust
xtreg IR BComp BComp_trend GDP INFL INFL_sq ALM U CR EXCH_rate ECB_rate
LITH_2014 covid_2020, re robust

// Hausman test
xtreg IR BComp BComp_trend GDP INFL INFL_sq ALM U CR EXCH_rate ECB_rate
LITH_2014 covid_2020, fe
estimates store fe_model
xtreg IR BComp BComp_trend GDP INFL INFL_sq ALM U CR EXCH_rate ECB_rate
LITH_2014 covid_2020, re
estimates store re_model
hausman fe_model re_model, sigmamore

// Housman Taylor
xthtaylor IR BComp BComp_trend GDP INFL INFL_sq ALM U EXCH_rate ECB_rate CR
LITH_2014 covid_2020 country_id, endog(GDP CR INFL INFL_sq EXCH_rate U BComp)
```